# Utilizing WaveFunctionCollapse Algorithm for Procedural Generation of Terrains using Remotely Sensed Elevation Data

Seyedparsa Dajkhosh
*department of computer science*
University of Tehran
Tehran, Iran
parsadaj@ut.ac.ir

*Abstract*—Procedural terrain generation plays a vital role in creating virtual landscapes for games, simulations, and various applications. The WaveFunctionCollapse (WFC) algorithm has proven effective in generating content by learning patterns from example data. In this research, we adapt WFC to generate terrain height maps using Shuttle Radar Topography Mission (SRTM) data. Instead of directly using raw height values, we use slopes to better capture structural features and preserve terrain patterns. Statistical comparisons, including histogram analysis, as well as evaluations of the mean, median, and standard deviation of input and output data, demonstrate that the algorithm effectively retains the input's structural characteristics while generating new terrain. the results show that WFC, with slope-based input and height-level transformations, can generate realistic terrain patterns for applications in game development and beyond.

*Keywords— Procedural Content Generation, Terrain Generation, Wave Function Collapse, Remotely-Sensed Data*

I. INTRODUCTION

Procedural terrain generation is a key tool in creating virtual landscapes for games, simulations, and other applications [5]. It helps in generating terrains without the need for manual design. Among the many techniques used in procedural content generation (PCG), the WaveFunctionCollapse (WFC) algorithm has gained attention. it can learn patterns from input data and generate new outputs that follow similar structures. This makes it a valuable method for generating both visual and structural content. In this work, we explore the use of WFC to generate terrain height maps. We use the Shuttle Radar Topography Mission (SRTM) data as input. Remotely sensed data such as SRTM is useful in PCG due to its ability to represent real-world.

One of the challenges in using height data directly in WFC is the large range of values, which can result in patterns that rarely match during generation. To address this, we use slopes—gradients in the x and y directions—as input instead of raw height values to better capture the structural features of the terrain. Our approach ensures the generated terrains preserve realistic patterns while overcoming challenges associated with using raw elevation data.

To discuss other works, Doran and Parberry introduced an algorithm that uses software agents to generate terrains. While this method eliminates the need for human intervention, the algorithm is overly complex and involves many complicated components [1]. Terrain generation using noise algorithms such as Diamond-Square, Perlin noise, and Value noise, as explained in [8], often produces repetitive results. Although these methods can create infinite terrain, making them appear more realistic still requires manual tuning of parameters, such as frequency and the number of octaves in Perlin Noise. For instance, the terrain generation algorithm used in Minecraft—arguably one of the most successful examples of terrain generation—relies on human experimentation and analysis to determine the best mapping of Perlin noise values to height at a given point [4]. More recently, deep learning approaches, such as those proposed by Ramos [7] and Tang [9], use various types of generative adversarial networks (GAN) to generate terrain. While these methods significantly reduce the need for human involvement, they demand high-end computational resources to operate effectively. The goal of this research is to overcome the challenges of the previous successful works while also generating realistic terrains by adapting WFC for terrain generation that leverages geospatial data and slope-based inputs.

II. DATA AND METHODS

*A. Theory of WaveFunctionCollapse*

WaveFunctionCollapse is a procedural content generation algorithm that uses local patterns from example data to generate new textures or structures. Initially developed by Merrel it gained recognition in the game development community years later when Gumin shared it on GitHub [3, 6]. WFC quickly became a popular method for generating textures and game levels due to its simplicity and effectiveness. The algorithm operates in three main stages: Extract, Observe, and Propagate.

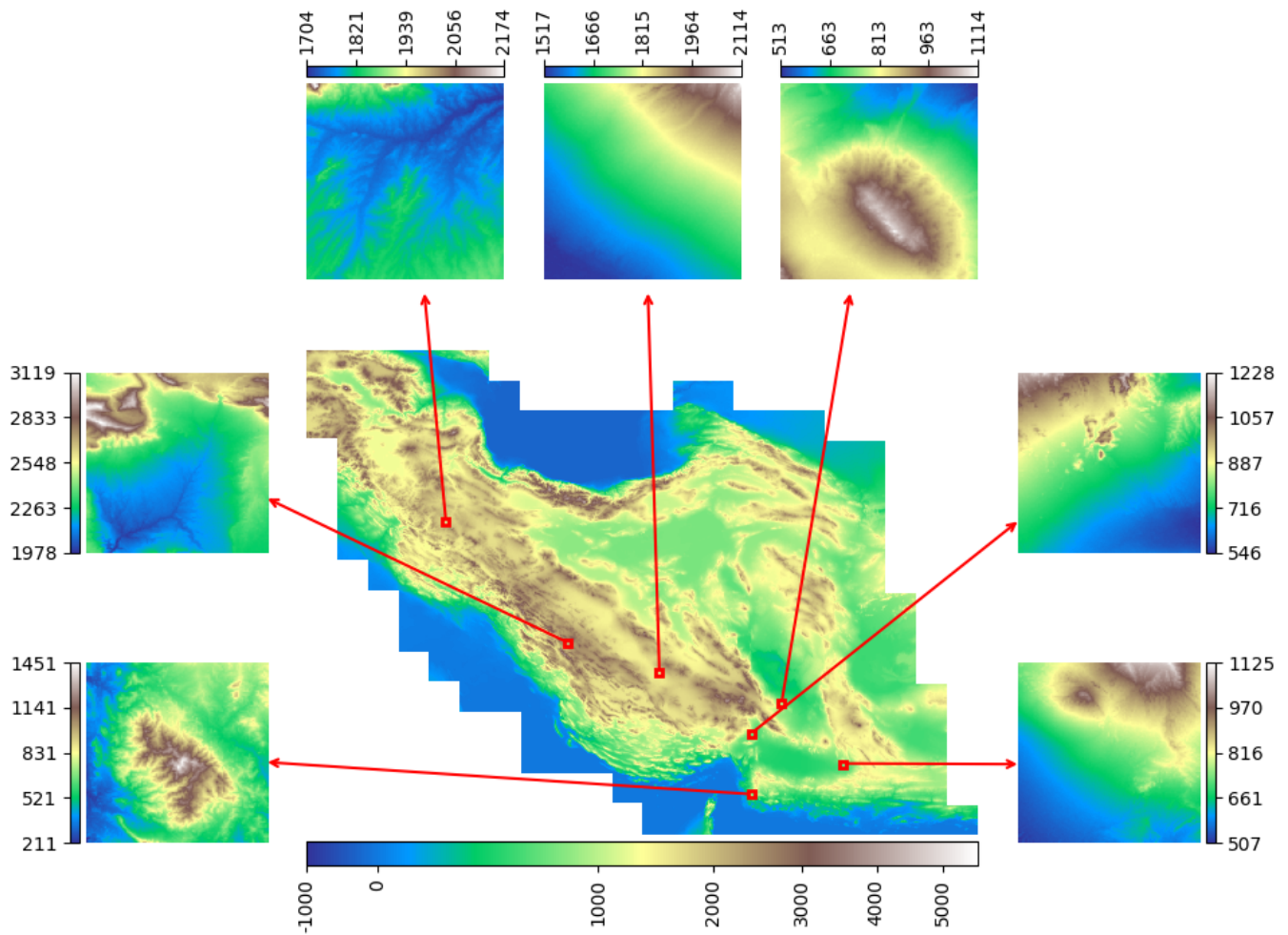

Fig. 1. The height map of Iran at the center, surrounded by the research areas.

In the extraction stage, WFC analyzes the input data to identify local patterns within a defined window size. These patterns represent small sections of the input, and their relative frequencies are stored for use in later stages. Along with the patterns, the algorithm records adjacency rules, which define which patterns can neighbor one another. One approach is to only use adjacency rules that directly appear in the original input by recording the occurrence of each pattern next to another. However, to improve effectiveness, one can iterate through all extracted patterns and determine additional valid adjacencies, ensuring they do not introduce contradictions. This step forms the learning phase of the algorithm.

The observation stage initializes a blank grid where each cell is said to be in superposition and can contain all possible patterns. To proceed, the algorithm selects a cell with the lowest entropy (which is highest constraints or fewest valid patterns in this research). Using the frequencies learned during extraction, it assigns a pattern to the selected cell probabilistically. Now we say that that cell collapsed into having only that possibility. This step moves the algorithm closer to generating a new content structure. After observing a cell, the algorithm updates its neighboring cells based on adjacency rules. Only patterns that match the assigned cell's constraints remain valid for its neighbors. This step propagates changes across the grid, reducing possibilities in nearby cells. If a contradiction arises (i.e., no valid pattern remains for a cell), the algorithm fails and restarts with new randomness. The cycle of observation and propagation continues until the grid is fully populated without contradictions [8].

*B. Data Acquisition and preparation*

The Shuttle Radar Topography Mission was launched by NASA in February 2000. Its goal was to map the Earth's surface by transmitting and receiving radar signals from a space shuttle orbiting the planet. The mission produced a digital elevation model with a spatial resolution of 1 arc-second (about 30 meters per pixel). This data is freely available on demand from various platforms [2]. For this research, data from seven locations in Iran were used. The names of the downloaded files can be seen in table 1, and Fig. 1 shows these areas of interest.

TABLE I. NAMES OF THE DOWNLOADED FILES

| N26E057 | N27E060 | N28E057 | N29E058 | N30E054 | N31E051 | N35E047 |
|---------|---------|---------|---------|---------|---------|---------|

a. Numbers in the file name indicate Easting (after E) and Northing (after N) of the image.

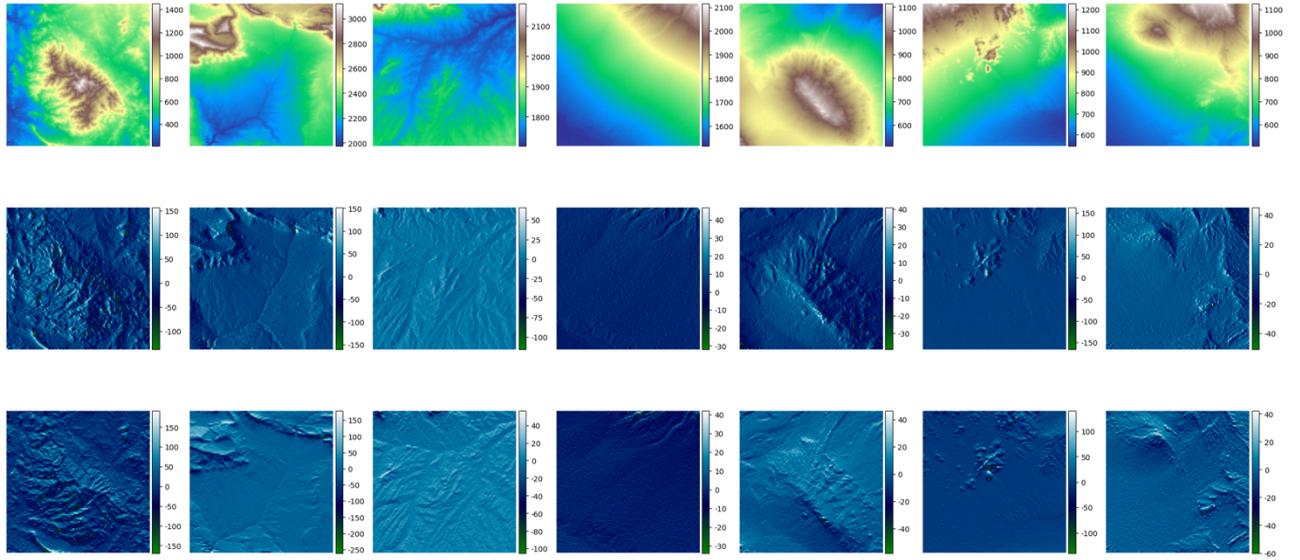

Fig. 2. Input data visualization. The first row displays height maps in meters relative to sea level, the second row represents the gradient in the x direction (Gx), and the third row represents the gradient in the y direction (Gy).

Since the downloaded images are large (3601 by 3601 pixels), a 100 by 100 window was selected from each region to reduce extraction time. Before this, the input image was down-sampled by a factor of 8 using bilinear interpolation. This step helped remove noise and smooth the input data.

The WFC works by extracting local neighborhood patterns from the input data. However, since heightmaps often contain many distinct values, running the algorithm directly on height data can lead to a large number of patterns with few neighboring matches. This can make the generation process less effective. To address this issue, gradients in the x and y directions are calculated and used as the algorithm's input instead of the raw heights. One can think this way, if all height values in a heightmap are increased by 1000, the terrain shape remains the same, but the extracted height patterns would differ significantly. In contrast, the slopes stay consistent, preserving the essential terrain features. The formulas for calculating the gradients are:

$$Gx[x,y] = H[x+1, y] - H[x,y] \qquad (1)$$

$$Gy[x,y] = H[x, y+1] - H[x,y] \qquad (2)$$

where Gy and Gx are the gradients and H represents height. Fig. 2 shows the calculated gradients in each input image. This results in input data of size 99 by 99 by 2, where the two channels represent the gradients in the x and y directions. Since only the valid gradients in the height image were calculated, the width and height of the slope images shape were reduced by one pixel.

To improve the chances of finding better local pattern adjacency rules, additional transformations can be applied to the input data. In this research, vertical and horizontal flips as well as 180-degree rotations were applied. These transformations help the algorithm identify more robust adjacency rules, enhancing the quality of the generated results. Unlike traditional WFC implementations that transform extracted patterns, our method avoids altering the gradients directly. Rotating or flipping gradient patterns would produce entirely new patterns, potentially losing the original structure of the terrain. By transforming the input raw heightmaps instead of the gradients, the integrity of the terrain's slope patterns is preserved. For example, flipping the heightmap ensures that both uphill and downhill slopes of the same magnitude are valid, effectively doubling the variety of patterns. These transformations increase the chances of finding more adjacency rules, making the generation process more robust and flexible.

## C. Slope Generation with WaveFunctionCollapse

### 1) Pattern Extraction

The WFC algorithm uses a 2 by 2 by 2 sliding window to extract patterns from the input slopes. This window moves across the entire grid, capturing all possible local configurations. Each pattern represents a small neighborhood of slope values. After extraction, the algorithm determines adjacency rules by comparing patterns. These rules determine how patterns can be placed next to each other. For two patterns to be adjacent, their overlapping rows or columns must match. For example, the bottom row of one pattern must be equal to the top row of another. Examples of extracted patterns and adjacency rules can be seen in Fig. 3 and Fig. 4. The chosen pattern size is a balance between computational efficiency and the ability to preserve terrain features. Larger patterns would result in fewer compatible adjacency rules, requiring more computational resources and larger input data. This would demand more computational resources and time, which is a limiting factor for this research.

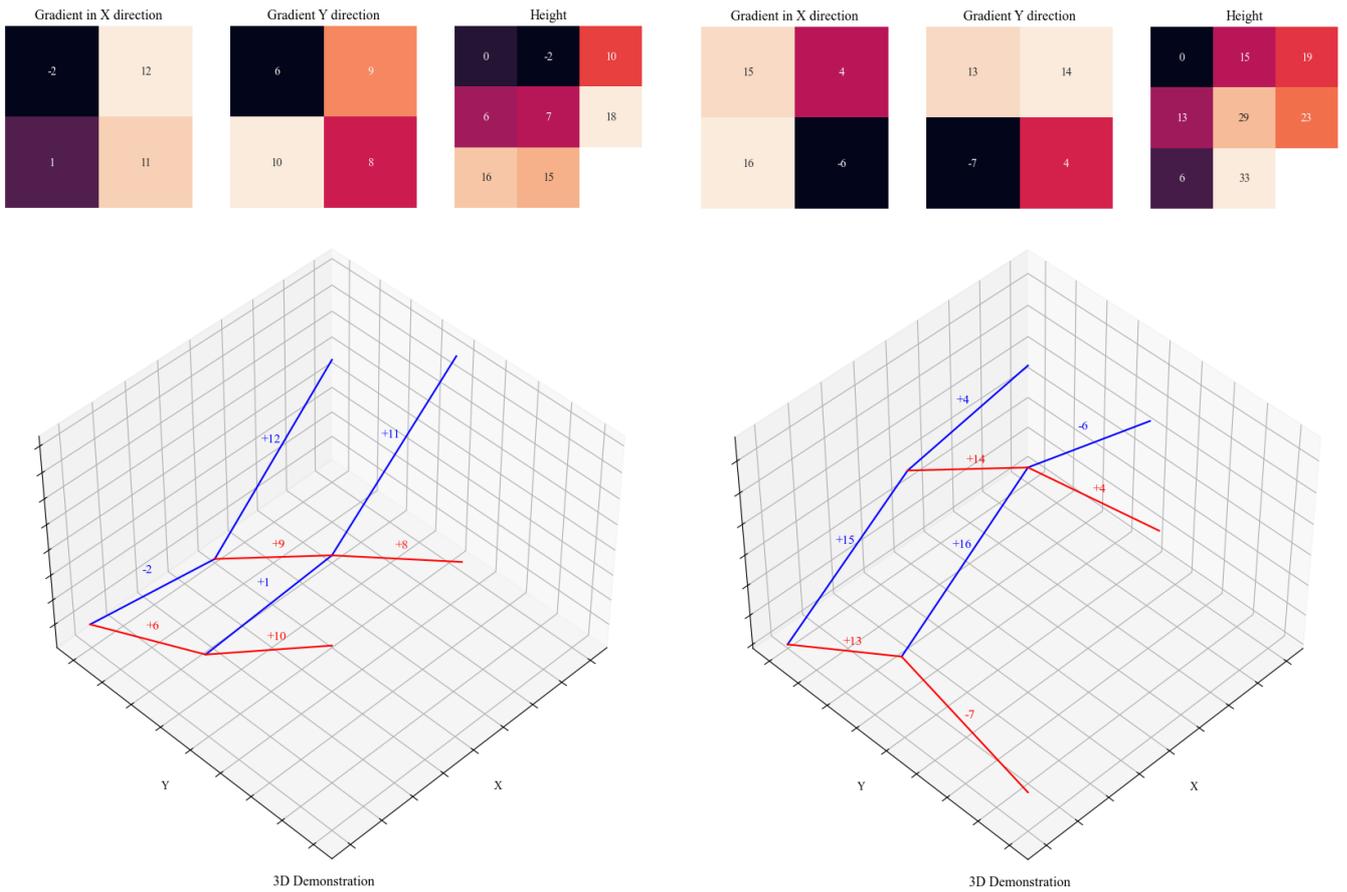

Fig. 3. Examples of extracted patterns. The top row displays the gradients in the Gx and Gy directions, alongside the reconstructed height map derived from these gradients. The second row provides a 3D visualization of the slope pattern for better spatial interpretation.

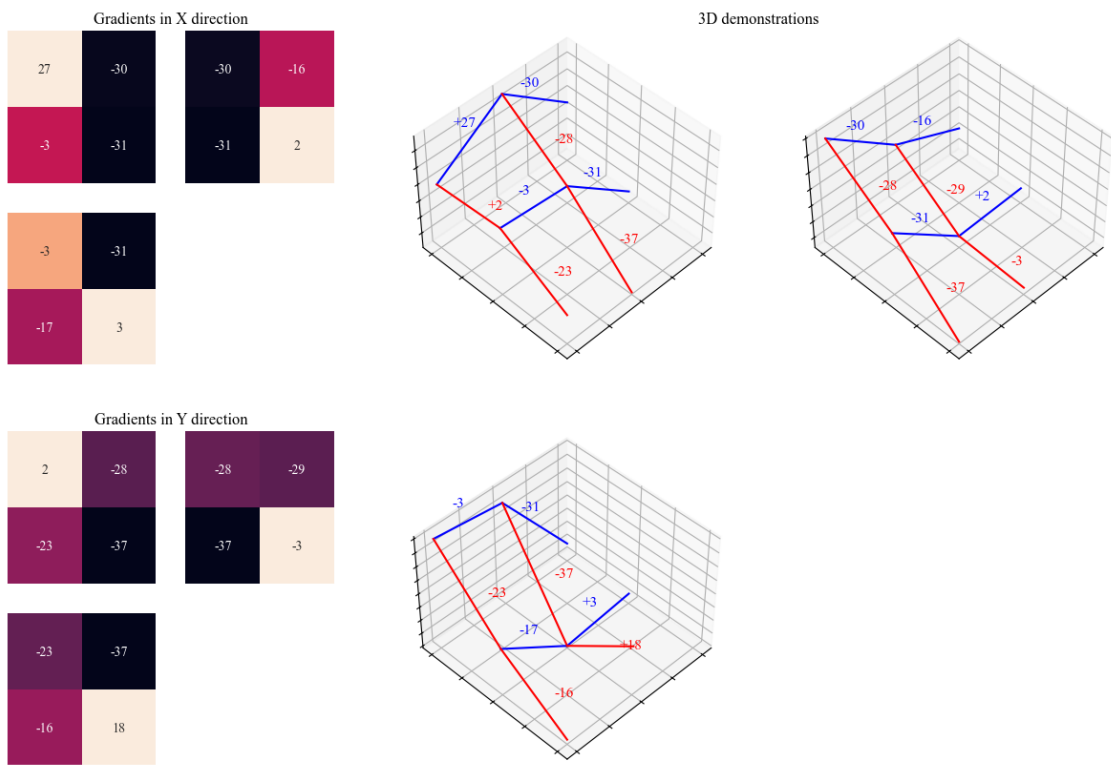

Fig. 4. An example of adjacency rules for one pattern with its right and bottom neighbors. The left side shows the slope patterns in the Gx and Gy directions, while the right side presents their 3D visualizations

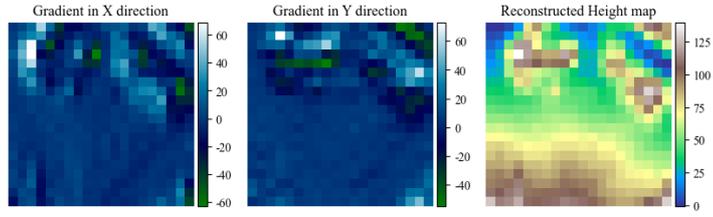

Fig. 5. generated gradients and their corresponding heigh map reconstruction

*2) Generating Outputs*

Using the learned patterns and adjacency rules, the algorithm can generate slope maps of any size by observing the cell with minimum entropy and propagating the constraint to its neighbors. These maps follow the structures and features of the original input. Once the slope maps are generated, they serve as the foundation for reconstructing height maps. A generated sample of gradients and their corresponding height map is shown in Fig. 5.

*D. Terrain Reconstruction*

The reconstruction process integrates the slopes generated by the WFC algorithm back into a height map using equation 3 and 4, reversing what was done in equation 1 and 2.

*1) Integration Along Rows:*

Starting with an initial height value, the algorithm uses the x-direction slopes (Gx) to calculate the height along each row (Hx). The height at position [x,y] is computed as:

$$Hx[x, y] = H[x-1, y] + Gx[x-1,y] \qquad (3)$$

*2) Integration Along Columns:*

Similarly, the y-direction slopes (Gy) are used to calculate the height along each column (Hy). The height at position [x,y] is computed as:

$$Hy[x, y] = H[x, y-1] + Gy[x,y-1] \qquad (4)$$

*3) Ensuring Consistency:*

Since the extraction used a 2 by 2 by 2 window, the integration process ensures that the calculated heights from both directions align consistently. This guarantees that for any point [x,y], the reconstructed height satisfies equation 5:

$$H[x, y] = Hx[x, y] = Hy[x, y] \qquad (5)$$

In Fig. 7, six input data and two generated terrains from each of them is shown.

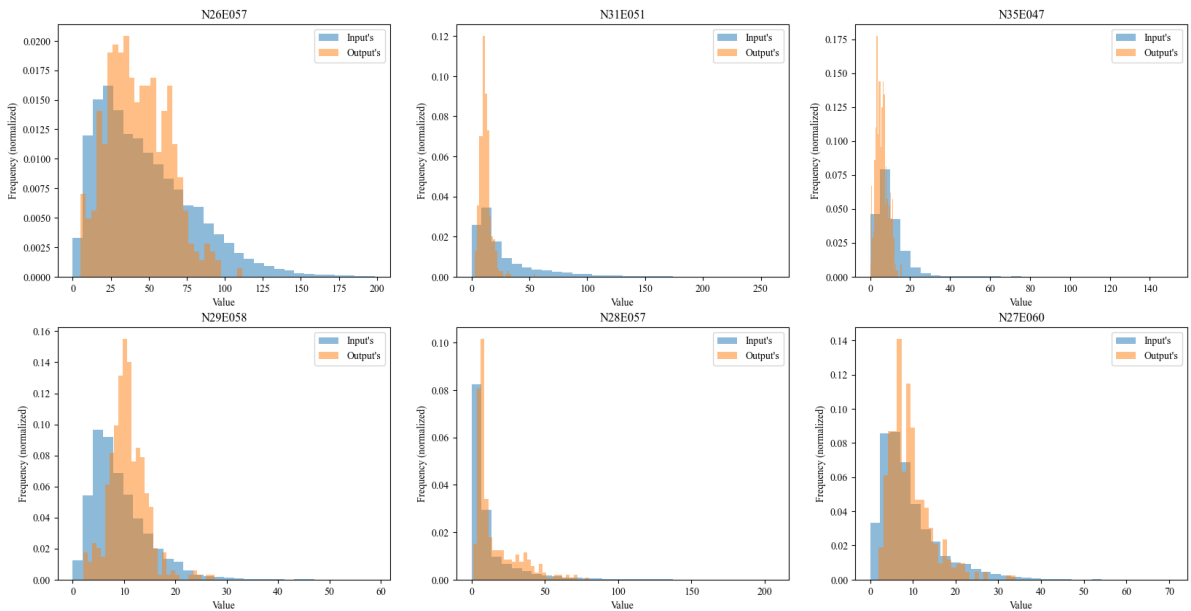

Fig. 6. histogram comparison of inputs and outputs

TABLE II.    STATISTICAL COMPARISON BETWEEN INPUT AND OUTPUT SLOPE MAGNITUDES

| File name | Std in | Std out | Mean in | Mean out | Median in | Median out |
|---|---|---|---|---|---|---|
| N26E057 | 31.3 | 19.7 | 47.7 | 43.1 | 41.1 | 41.6 |
| N31E051 | 28.1 | 5.1 | 27.1 | 11.9 | 16.5 | 11.2 |
| N35E047 | 7.4 | 2.9 | 9.8 | 5.9 | 8.2 | 5.4 |
| N29E058 | 5.8 | 3.6 | 9.2 | 10.7 | 7.8 | 10.3 |
| N28E057 | 17.3 | 14.8 | 12.4 | 15.9 | 6.0 | 8.4 |
| N27E060 | 7.0 | 4.8 | 9.2 | 9.4 | 7.1 | 8.5 |

## III. Results

The algorithm successfully generated slope patterns for 6 of the 7 input images and then These slopes were reconstructed into height maps. However, for one input image, the algorithm was unable to generate valid outputs despite multiple attempts, indicating a potential limitation in pattern compatibility or diversity for that specific input.

Fig. 6 shows histogram plots comparing the frequency of slope magnitudes in the input and output images. Both histograms display similar distributions for each input-output pair, indicating that the algorithm preserves the statistical characteristics of the input slopes during generation. Table 2 provides additional statistical parameters, which are mean, standard deviation, and median in meters per pixel, for both the input and output slopes. The values in the table, further confirm the similarity between input and output, suggesting that the generated terrains maintain the key features of the original data.

## IV. Discussion

### A. Strenghts of the Approach

The main advantage of this approach is that it doesn't require parameter tuning, as it only needs the input data to generate the output. Unlike methods like Perlin noise, which need a lot of parameter adjustments, this method simplifies the process and still produces realistic results. Also, Using slopes instead of raw height maps in the WaveFunctionCollapse algorithm worked pretty well and by transforming the data at the height map level, rather than modifying the extracted patterns, the integrity of slope patterns was maintained.

### B. Challenges and Limitations

- This approach has computational limitations, particularly with larger pattern sizes or higher-resolution terrains. For instance, in this research, extracting adjacency rules for a 100x100 image took about three hours, and this time would increase nonlinearly for larger inputs. Additionally, adjacency rules sometimes resulted in repeated patterns, particularly when a pattern was a neighbor to itself. Although this issue occurred during experiments, the specific data was lost due to a bug in saving the algorithm's state.

- There is a high probability of failure in the generation process. In this research, aside from the one image where no solution was found after dozens of attempts, even in successful generations, only one in every ten tries would result in an output. When the algorithm failed, it had to start over from the beginning to try generating a new solution.

- When input images are excessively large (e. g. Using all of the SRTM images), the adjacency rules may become too broad and meaningless. This could result in outputs resembling random slope generation rather than structured terrain synthesis, as the patterns might be neighbors with many other patterns containing diverse values. This randomness reduces the effectiveness of the algorithm in generating realistic terrain.

### C. Future Work

To address the challenge of excessive patterns in large training datasets, a hybrid approach combining WFC with a Markov chain could be explored. In this method, an initial map with limited slope categories (e.g., flat, low, medium, steep) would be generated using WFC. Then, specific slope values could be determined based on the pattern frequencies corresponding to these categories. This two-step process could help balance diversity with manageability, ensuring more structured and realistic terrain outputs. This approach would also enable the use of larger pattern sizes and input sizes. By limiting the possible outputs to a smaller set of categories, the number of patterns would be significantly reduced. As a result, the extraction time for adjacency rules would be greatly decreased, making it more feasible to handle larger datasets and more complex terrain structures.

Another possible direction for future work is to develop a method to handle failure attempts during the generation process, rather than requiring the algorithm to start over from scratch. This could involve implementing a strategy that allows the algorithm to backtrack or adjust its approach when it encounters a failure, thereby improving efficiency and reducing the need for repetitive attempts.

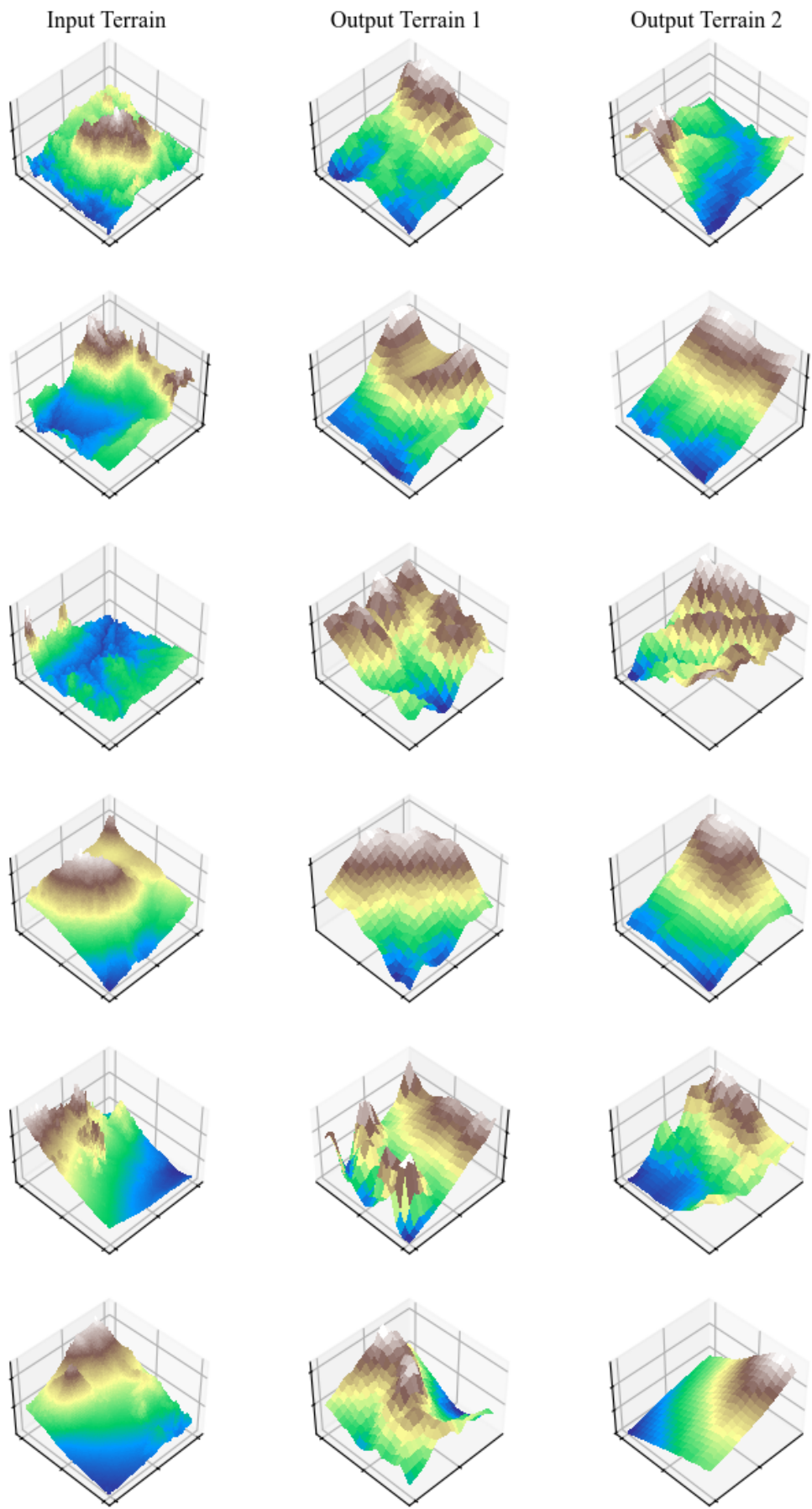

Fig. 7. Outputs: Each row displays the input (left) and two corresponding outputs (center and right).